# BROWNIAN MOTION, IONIC FLUX, CATALYTIC REACTION AND HETEROGENEOUS NUCLEATION IN BIOLOGICAL SYSTEMS

P. C. T. D'Ajello and P. R. Hauser

Departamento de Física UFSC/CFM
P. O. Box 476 – CEP 88040-900
E-mail:pcesar@fisica.ufsc.br
Fax:++55 231 9688

**ABSTRACT**

A model to describe the arising of new structures in an initial homogeneous biological system is proposed. The essay is motivated by the intention to work on a non-equilibrium situation grouping together several mechanisms and processes as: catalytic reactions on a surface, diffusion, stimulated migration and selective heterogeneous reaction. A model for morphogenesis in early embryos is developed on two basic assumptions; (i) the existence of an electrified surface that defines the shape (form) of the growing structure and (ii) a mechanism to select morphogens (ions or free radicals) from an initially homogeneous medium. The homogeneity is broke when an electric potential arise between different parcels of the system, triggering a complex dynamic that drive the development of material deposits into localized regions of the space. The evolution of the deposits is described by a stochastic formalism allowing for analytical expressions relating macroscopic.

# I. INTRODUCTION

Any living organism need, for its sustenance, to keep up an open interaction with the environment. This interaction could refer to a relation between the entire organism and the external world, or to a connection among internal subsystems. Any way, to ensure survival, optimizing chances, a regulatory flux mechanism has fundamental importance. The importance of mass transport processes is evident in sophisticated organic species that maintain a constant functioning to select composites, from the physiological solution. This selectiveness in high-organized systems is related to a flow control ability. This control has a central importance during early embryos development when reactions, at specific locations, must occurs during specific time intervals. However, a naive description of ontogenesis compels to consider (i) cellular differentiation, (ii) the increase in the germ mass (biological growth) and finally, (iii) morphogenesis during embryo's development. This reference is enough to point that a growth processes involves a few more things than just an adequate description of the flow of matter. One of the basic questions is connected to the onset of new structures, form and composition, from an initially homogeneous medium. How to get a liver or a rib from the same physiological solution? The diversity emerging from a single background emphasizes that nucleation and growth of new structures is much more dependent from the dynamics that generate the fluxes than properly by the flux itself. To conceive a mechanism to compute the overall results in a system with competitive dynamics is the goal of those who try to describe any growth process.

It has been remarked that in living organisms the physiological solutions are mainly composed of high reactive ionic elements or organic radicals then, it is natural to conceive matter transport inside biological cavities as a very complex process, guided by the competition of many mechanisms. Nowadays a lot of inspiring works focus attention to one or another aspect involved in this complexity. It has been pointed out that concentration gradients (Wolpert 1981; Gierer 1981), diffusion (Crick 1970), selective chemical reactions (Babloyantz 1977) and ionic currents (Stern 1986 and Jaffe 1981) are some of the effects invoked to maintain morphogen's fluxes.

In spite of the wonderful analysis found in the literature there are two basic problems that claim for adequate statement when a synthesis of the subject is intended. First is (i) the allegation about the source of the instruction to break symmetry in order to displace systems from equilibrium states with homogeneous mass distribution. A necessary condition to promote morphological alterations. Second, there is a technical problem. (ii) To work complex systems in order to get plain and causal laws that explains the growth processes. Although these questions demands an effort we are still unable to do, it is time for an endeavor to search a scheme where different mechanisms, working together, generates a tractable non equilibrium dynamic that describes the onset of new structures in biological systems. The articulation of these mechanisms in just one simple theory still lacks, however it seems that electrochemical transport mechanism could be an adequate proposition to be explored. In fact through an electrochemical description we integrate generalized forces like the concentration gradient and electric forces on a description of the mass/charge flow near catalytic surfaces. In addition it is well known that electrochemical processes manifest the ability to induce selective migrations through subtle changes on the electrode electric potential.

In this paper we propose a model to examine biological systems in transient process. Our intention is founded on a stochastic process that group together: catalytic reactions on a surface, diffusion, stimulated migration and selective heterogeneous. The model offers specific predictions about the ionic currents that flow through membranes such its results can be compared to experimental verifications. Following Turing (1952), it is convenient to advise that the model is a simplification and an idealization. It recognize the cell theory but idealize it into continuous space where some points are associated with particular cells. The organic material is also continuously distributed inside the system that is visualized as an electrolytic cell, fulfilled by a physiological solution composed by different chemical species diluted in water. Among the composites a particular ionic specie (we call morphogen from now on) is investigated. This "particle" is supposed to perform a free Brownian motion into the bulk of the solution. Closer to a singular plane of cells the morphogen becomes aware of a field which, modify its chemical nature and its motion. The change in morphogen's dynamics follows a particular reaction which take places on a limited region of the space. After this change the morphogen perform no more a free Brownian motion but a stimulated migration, being addressed to the electrified plane to react on particular points. Through these points the "particles" are absorbed by the cells such a deposit grows by the successive arose of an specific type of particles. The cells, stuffed by the new material, are supposed to be still alive, remaining with an integral semi-permeable membrane. That is, we disregard the conception of a close packing material inside the cells. To illustrate the approach we think on the selective absorption of Calcium to originate a bony structure. In this picture, Calcium ions play the role of a morphogen.

On what follow we analyze the motion of just one morphogen, selected among the others by the value assumed by the electrical potential difference of the cellular surface. We consider the flux of ions through a collection of points on a plane of cells. These points are called nucleation points in analogy to early electrochemical guiding studies (D'Ajello et al. 1999 and D'Ajello et al. submitted to publication). They indicates the loci of points, on through which, the morphogens are ingested by the cells.

In section II the model is defined and the procedures to achieve the mathematical solutions are indicated. In section III two particular situations are analyzed, namely; (a) one in which diffusion is the dominant mechanism to define the dynamics of the morphogen in its flight to the plane and, (b) the other in which the chemical reaction rate of the morphogen, on the cellular surface, is the slowest process. Some results are sketched and discussed in the fourth paragraph. The conclusions are at the end.

## II. THE THEORETICAL MODEL.

Biological living organisms are the prototype of non-equilibrium thermodynamic systems and the existence of time dependent processes in heterogeneous environment is its preeminent characteristics. Notwithstanding, biological systems are well regulated in every one of their parcels to the point that severe changes, occurring in short elapsed time, are frequently an evidence of a disease or a high risk integrity process. So, to work with health situation, we assume that our system is too short compared to the entire organism such the concentration gradients inside changes very little in distances less than 2000 $\mu m$ (a distance defined by some 200 cell). That mean, we choose to work with a medium where a great number of particles are diluted in a physiological solution, mainly constituted by water molecules. The assembly particles form a set of many species having different mass, different constitution (they could be single ions or complex molecules) and different ionic states. Among the myriad of species, we focus attention in one particular kind of particles what is devised as an hydrated/solvated ion showing a homogeneous distribution into the bulk of the system. To simplify we call this singular ion a morphogen. The morphogen encapsulated by a shell of water molecules, due to polar attraction, moves inside the electrolytic medium performing a Brownian motion. This complex homogeneous medium undergo a symmetry break at time, $t=0$, when an electric potential difference is setting up. It is not relevant, at this moment, to discuss the mechanism that triggers sustain this electric potential but once the symmetry is broken, a biological surface is precisely conformed. Instead to work with the entire surface we choose to consider a limited portion of it assuming that every event in this sub-system occurs in a similar way as on the integral surface. Thus we look to plane of cells, that works as an electrode in an electrochemical cell. There is always a second electrode (the positive electrode), but far enough from our system such it is natural to disregard. On the electrode (the plane of cells), there are many electroactive points (we call them, nucleus, from now on). Through these points the cellular surface perform a charge exchange with the morphogen. Some of these points are automatically activate at $t=0$; others turn to be activated during the course of time, depending on the physical and chemical conditions imposed to the system. In Fig. 1a we show a schematic representation of this conception where. A section of the plane is visualized at two successive instant of time. In this transversal section of the plane the right-hand column of cells correspond to topmost shell of the electrified surface. On this shell are located the nucleus or the source of charges. It is also through these nucleus that morphogen are absorbed by the surface to a posterior diffusion or sedimentation.

We are advised that a little change in the magnitude of the electric potential (EP) is a sufficient conditions to select another kind of particle to be attracted in order to overcome a reaction on the catalytic plane. Thus the magnitude of the electric potential select the type of morphogen to be activate to react on the plane. In conclusion, the morphogen is one specie of the many solvated ions inside the physiological solution and, its choice is dictate by the magnitude of the EP.

In the vicinity of the electrified plane our system is very complex, clearly out of equilibrium and affected by many different forces as: electrical, gravitational, buoyancy and some generalized thermodynamic forces. From the competition of these many effects arise the overall dynamics followed by the morphogen.

At large distances (greater that some cents of micron) from the plane/electrode the morphogen are assumed as free of forces and evolve describing a Brownian motion. Closer to the plane, the morphogen feels the effect of a field that change its electrochemical character (it loose it solvatation shell). Once the ion is stripped of the hydrated molecules, it turns to be very sensitive to the field generated by the electrified plane. In this charged state, the morphogen migrates to the plane to be reduced/oxidized. In a neutral state it is absorbed in special points (the nuclei). The continuous absorption of these particles gives place to a material deposit inside the most external cells that conforms the surface. The superposition of deposits, arising from neighborhood nuclei, develops a three-dimensional structure whose form depends on the geometry of the entire electrified surface as well as of the spatial nucleus distribution on it. For simplicity we work with nuclei disposed on a plane of cell but it is immaterial if they are disposed on any other kind of surface to form different morphological structures. A natural extension of this approach consist to consider an electrified surface where the potential in no more constant but change periodically along its extension such some ensemble of nucleus react with one particular kind of morphogen whereas another set of nucleus attract a different type of them. Thus two or more distinct particles can be simultaneously absorbed by the surface. This is an appealing idea because it denote the possibility to get different compositional structures inside the volume defined by the electrified surface without discarding possible reactions among the absorbed morphogens. This hypothesis favor diversity and plasticity in living systems. Here we limit ourselves to the simplest case, schematized by the four steps listed below.

1. The morphogen, with a solvatation (hydration) sphere, migrate inside the solution bulk describing a free Brownian motion.
2. Morphogen, which randomly cross a hemispherical surface, defined by a radius $R$, will loose their solvatation shells. As shown in Figure 1b this surface will be named first reaction surface and the radius $R$, reaction radius (a critical distance around a nucleus). The rate at which the removal of the solvatation shells occurs on the surface defined by the radius $R$ changes according the physical situation. The radius $R$ will be assumed constant during the growth process.

3. Inside the first reaction surface (region II in Figure 1b) the morphogen migrate towards an electro-active site performing a diffusive motion assisted by an electric field.
4. At the electro-active site, morphogen are reduced and ingested to improve the growth. Furthermore, crossing the first hemisphere is not a necessary and sufficient condition for the occurrence of a reduction/oxidation reaction and consequent aggregation of the reduced/oxidized morphogen to the growing nucleus.

As a preliminary to the analytical development of the model, some considerations are necessary.

a) It is assumed a random distribution of nucleus on the plan of cells.

b) From a theoretical point of view, R is introduced to mimic a cut off of some potential or generalized thermodynamic potential. That is, R is considered as a mean distance, from a nucleus, defining a region where the effect of some field, acting on the reacting species, no longer can be ignored. Complexity is always connected to systems where many variables, parameters and processes can be present. Frequently, the success of a particular description rests in the ability to choice a few numbers of them, the most important ones. In our case, R plays an important role demarcating the interface between two regions where distinct mechanisms dominate. The introduction of this parameter turns possible to obtain an analytic expression for a nonequilibrium process where spatial inhomogeneities are taking into account.. However, R is not a universal constant and its magnitude changes according the physical situation without affecting the model.

c) To solve the problem in a realistic way, it is necessary to choose a physical macroscopic variable that could be compared to those arising from real the experiments. Having conceived the morphogen as an ion, it seems natural to use the electric current density, flowing through the plan of cells during the electro deposition/absorption process, as the measurable physics parameter.

d) Once the motion of a morphogen is Brownian, the analysis must be carried in a probabilistic manner therefore, one of our goals is to determine the probability to found a morphogen a distance $R$ apart from the center of a super-particle (the nucleus). By this procedure we obtain the flux of the morphogen through the surface of the first hemisphere. Then, knowing the probability that a morphogen is on a nucleus, after to enter inside the first reaction hemisphere, we determine the current of the incoming material.

## III – FORMAL MATHEMATICAL PROCEDURES

Following the schema devised in Fig. 1, we seek for a solution that satisfy different boundary conditions in a partitioned space. We work adopting a two step procedure. First we derive the flux probability, $j_h(R,t)$, that describe the rate the morphogen cross the first hemispherical surface and, then, we define the probability, $p_d(r',t)$, that a morphogen was on a nucleus following its incoming in region II as predicted before. The current density corresponding to absorbed morphogen on the plan or, what is equivalent, the growth rate of the new structure is:

$$I = N.\,p_d(r',t).\,j_h(R,t)\,. \qquad (1)$$

$I$ is the measured current and, $N$ the number of nuclei on the plane of cells.

### $j_h(R,t)$ derivation

As stated before, inside region I the solvated morphogen perform a free Brownian motion. On the surface of the first hemisphere they react loosing their solvated shell. This diffusion-reaction process contemplates two basic situations characterized by the relative magnitude of the diffusion constant, $D$, and the reacting rate, $K$ (the rate the morphogen loose their hydration shell). (a) when $K \to \infty$, diffusion is the slowest process and the dynamic is controlled by mass transport. (b) when, diffusion turns to be low, if compared to a $K = finite\ \ value$, the reaction on the surface of the first hemisphere is the slowest process and the dynamic is regulate by reaction.

In both cases, to obtain the probability densities that a Brownian particle is on the first reaction hemisphere we use the Langevin formalism (Nicolis & Prigogine 1977) represented by the equations:

$$\frac{\partial r}{\partial t} = v, \qquad (2)$$

and

$$\frac{\partial v(t)}{\partial t} = -\beta v(t) + A(t) + F_E\,. \qquad (3)$$

In Eqs. (2) and (3), $r(t)$ and $v(t)$ are, respectively, the position and velocity of the Brownian particles (hydrated morphogen), $\beta$ is the usual viscosity divided by the particle mass; $A(t)$ is the stochastic force characterized by a zero mean value and a delta correlation. $F_E$ is the external deterministic force. Assuming $\beta$ sufficient large that velocity is heavy damped respecting $r(t)$, $v(t)$ is eliminated by an adiabatic operation that simplify Eq. (3). The associated Fokker-Planck (FP) equation (when $F_E = 0$) in spherical coordinate is,

$$\frac{\partial\left(rp(r,t)\right)}{\partial t} = D\frac{\partial^2}{\partial r^2}\left(rp(r,t)\right). \qquad (4)$$

The diffusion coefficient $D$ is related to the second moment of the distribution.

It is important to note that the probability density can be related to the concentration of the Brownian particles through the expression,

$$p(r,t) = \frac{c(r,t)}{c_o}, \qquad (5)$$

where $c(r,t)$ is the morphogen concentration at a distance $r$ from the nucleus and $c_o$, its concentration under equilibrium conditions at time $t = 0$.

Equation (4) is solved obeying two distinct set of boundary conditions, corresponding to the two situations describe before, namely:

**(a)** $K = \infty$ **(reaction control)**

With the boundary conditions:

$$p(r,0) = 1 \qquad \forall r \rangle R \quad , \qquad (6)$$

$$p(r,t) = 0 \qquad r = R, \quad \forall t \quad , \qquad (7)$$

Relation (7) mimic the fact that an hydrated morphogen is transformed by reaction on the first reaction hemisphere. Using Fourier transform techniques eq.(4) gives the solution,

$$p(r,t) = 1 - \frac{R}{r} + \frac{R}{r}erf\left(\frac{r-R}{\sqrt{4Dt}}\right). \qquad (8)$$

**(b)** $K =$ **finite value (kinetic controlled reaction)**

In this case the boundary conditions are,

$$Kp(R,t) = 2\pi R^2 D\frac{\partial\, p(r,t)}{\partial r}\Big|_{r=R}, \qquad (9a)$$

$$p(\infty,t) = 1 \qquad \forall t \quad , \qquad (9b)$$

and

$$p(r,0) = 1 \qquad \forall \quad r\rangle R. \qquad (9c)$$

Condition (9a) expresses the possibility that the hydrated morphogen have a non-null probability to avoid the reaction/transformation on $R$. That is, only a fraction of the Brownian particles, crossing the hemispherical surface, gets a transformation. Condition (9b) states the diffusion-reaction process is effective only within finite distances from the reacting surface. Far from the surface, the distribution of hydrated morphogen remains essentially the same as in equilibrium. Finally, condition (9c) defines the initial condition at any distance $r$ ($r \rangle R$).

Under these conditions Equation (4) gives the solution,

$$p(r,t) = 1 - \frac{(R-\gamma)}{r}\left\{ erfc\left[\frac{r-R}{\sqrt{4Dt}}\right] - \exp\left(\frac{Dt}{\gamma^2} + \frac{r-R}{\gamma}\right) erfc\left[\frac{\sqrt{Dt}}{\gamma} + \frac{r-R}{\sqrt{4Dt}}\right]\right\}, \qquad (10)$$

where

$$erfc(x) = \frac{2}{\sqrt{\pi}}\int_x^{\infty} \exp(-z^2)dz \,, \qquad (11)$$

and

$$\gamma = \frac{1}{\left(K/2\pi DR^2\right) + \left(1/R\right)} \,. \qquad (12)$$

Given the probability densities (Eqs. (8) and (10)), it is possible to evaluate the probability flux of morphogen across the first reaction hemisphere when it loss its hydrated shell,

$$j_h = 2\pi R^2 D \frac{\partial p_R(r,t)}{\partial r}\Big|_{r=R} \,,$$

The current density across the hemispherical surface is:

$$j_h = 2\pi RzFDc_o\left(1 + \frac{R}{\sqrt{\pi Dt}}\right), \qquad (13)$$

when the reaction rate, $K$, goes to infinite, and:

$$j_h = c_o KzF\left[\frac{\gamma}{R} + \left(1 - \frac{\gamma}{R}\right).\exp\left(\frac{Dt}{\gamma^2}\right).erfc\left(\frac{\sqrt{Dt}}{\gamma}\right)\right], \qquad (14)$$

if reaction occurs at a finite rate on the first reaction hemisphere. In Eqs. (13) and (14) $z$ stands for the charge number characteristic of an ion (morphogen without its hydrated shell) and $F$ is the Faraday constant. Thus $j_h$ have the dimension of a electrical current density.

**Evaluation of $p_d(r',t)$ and morphogen flux to new structures**

Instead to use again a Langevin type formalism to describe the dynamic followed by the morphogen inside region II a naive scheme is adopted to calculate $p_d(r',t)$. $p_d(r',t)$ is the probabilistic weight expressing the fraction of morphogen that arrive at the a nucleus after its former reaction on the first hemispherical surface. Thus $r'$ represents a position inside region II.

To circumvent a formal calculation of $p_d(r',t)$ we observe that this probability must be proportional to the transformed volume occupied by the deposit, that is, the volume occupied by morphogen absorbed by a nucleus. Once the reduction on the nucleus can be conceived as a heterogeneous catalytic reaction use a know expression that give the time evolution of the volume occupied by the reacting particles per unit volume of space:

$$\frac{V_{tr}}{V_o} = \frac{transformed \quad volume}{unit \quad volume \quad of \quad space} \approx 1 - \exp\left(-\left[n_o kt + \alpha kt^2\right]\right). \qquad (15)$$

This expression describes the growth of deposits in kinetic phase transformation and it was derived in by D'Ajello, Kipervaser et al. (submitted to publication) as an equivalent version of the Avrami's theorem (Avrami 1939-41). In Eq. (15) $n_o$, $\alpha$ and $k$ are constants. They represent the number of active nuclei at $t = 0$, the rate of nuclei activation and the morphogen reaction rate (reduction) on the electrified surface respectively.

The final expression for $p_d(r',t)$ is obtained assuming its proportionality with the volume fraction, i.e.:

$$p_d(r',t) = b.\frac{V_{tr}}{V_o} = b.\left\{1 - \exp\left(-\left[n_o kt + \alpha kt^2\right]\right)\right\}, \qquad (16)$$

where $b$ is an appropriated constant factor.

By use of Eq. (16) and Eq. (13) we obtain the current density, associated to the morphogenetic flux through the plane of cells,

$$I = N.b.C_1.\left(1 + \frac{R}{\sqrt{\pi Dt}}\right).\left(1 - \exp\left(-\left[n_o kt + \alpha kt^2\right]\right)\right), \qquad (17)$$

when the strip reaction on the first reaction hemisphere is infinitely fast. Using Eq.(16) and Eq.(14) the current transient is given by,

$$I = N.b.C_2.\left[\frac{\gamma}{R} + \left(1 - \frac{\gamma}{R}\right).\exp\left(\frac{Dt}{\gamma^2}\right).erfc\left(\frac{\sqrt{Dt}}{\gamma}\right)\right].\left(1 - \exp\left(-\left[n_o kt + \alpha kt^2\right]\right)\right), \qquad (18)$$

but in this case the strip reaction, on the first reaction hemisphere, has a finite value.

## IV – RESULTS AND DISCUSSION

Equations (17) and (18) gives the current associated to the reduction of morphogen incorporated in the cells. The magnitude of the surface is not relevant but its electric potential and the number of nuclei per unit area it contains. In figure 2 we shown the current density curves generated from Eq. (17) to an arbitrary case where: $D = 1.x10^{-5} cm^2 / s$, $N = 1.x10^7 1/cm^2$, $n_0 k = 1256{,}6 \ 1/s$, $\alpha = 0$. In this case the free parameter is the reaction radius, $R$, to which is assigned different values as indicated in the figure caption. In all curves $K \to \infty$ on the first hemisphere, such diffusivity has fundamental importance to regulate the process because it is always the slower variable that regulates the dynamic. All current curves shows a peak, at the very beginning, followed by a retraction on its intensity toward a stationary value. According the magnitude of the parameter $R$ faster is attained this stationary value for the currents. The theoretical curves of Fig. 2 also shows that diminishing $R$ one soften the maximum and soon the current are stabilized. Due to a lack of experimental result we examine this assertion by inspection to electrochemical realizations and a good correspondence is found, see D'Ajello, Fiori et al. (submitted to publication). Notwithstanding the differences between an electrochemical cell and a biological membrane some typical behaviors must remain unchanged. For example, in electrochemical experiments an increase on $R$, corresponds to an increase on the magnitude of the electrostatic potential. From this we infer that and increment on the electrostatic potential abbreviate the transient regime because it is increased the driven force acting on the particles. In Fig. 3, a set of curves, obtained from Eq. (18) is depicted. They also correspond to current transient but, in this case the dominant mechanism is the chemical reaction rate on the first reaction hemisphere (quantified by $K$). In Fig. 3; $D$, $N$, $n_0 k$ and $\alpha$, are the same as in Fig. 2 but $K$ run through two orders of magnitude whereas $R$ is fixed at $3.16x10^{-4} cm$. The curves show that a decrease in $K$ is correlated to a reduction of the elapsed time necessary to reach the stationary value. In some sense a reduction in $K$ is equivalent to a reduction in $R$ when working the expression (17). A comparison with electrochemical realization allows to infer that a diminution of $K$ corresponds to an increase on the electrical conductivity of the medium. This increase on the electrical conductivity of the medium, by they turn, favor the mass transport and reaction kinetic tend to prevail, over diffusion, to regulate the dynamics. The similar behavior sketched by the curves in the cases, $K \to \infty$ and $K \langle\langle 1$, could gives the erroneous impression that just one expression is necessary to relate both cases. That is not true. To verify this limitation we included the curve (e) in Fig. 3. This curve emphasize a pathological behavior of expression (18) when $K$ is greater than a critical value.

In Figs. 4 and 5 we show the same results as in Figs. 3 and 4 the difference rests on the double normalization procedure which normalize the current density, $I$ and the time, $t$. In Fig. 5 two experimental curves, indicated by full circles and full squares, are introduced. They show the current transients obtained during metallic deposition on n-type silicon inside an electrolytic cell. They stress the fact that a little potentiostatic difference originates very disctint deposition rates. The experimental curves are obtained in situations where the electric potential difference between the deposition plane and the reference electrode indicates $-1300\,mV$ (full circles) and $-1100\,mV$ (full squares).

The normalized graphics could be compared to make evident that, under reaction control ($K$ smaller than $D/R^2$); a little time interval is spent to achieves the stationary value for the current signal.

It is important to observe that the current density curves never goes to zero. This fact, connected to biological systems, could be associated to a constant renovation of the cells inside an organic structure.

By economy we not shown a situation where $\alpha \neq 0$. If this parameter differs from zero there is a continuos increment on the nuclei density and the main consequence is to change the derivative of the current curves in the interval $0 \leq t/t_{max} \leq 1$.

Up to now no explicit function has been proposed to relate the currents with the electric potential, however D'Ajello et al. 1999 and D'Ajello, Fiori et al. had argued the this relation does exist through the reaction radius.

In our essay we have considered current curves generated at fixed values for the potential, that is, only one type of morphogen is reduced at the surface. An improvement consists to consider the existence spatial dependent potential such many different kinds of morphogen can be absorbed at the same time following the magnitude of the electric potential in each point on the surface. This assumption encourages the existence of many reactions inside the electrified surface, involving many types of morphogens.

## V – CONCLUSIONS

The expressions for the current transient occurring in developmental processes are introduced. Two limiting situations have been examined to indicate that ionic flow differs during the early stages of the process. This differences are connected to the many mechanism that regulates the dynamic. On the phenomenological point of view the model offers many possibilities, it favor an alternative to understand the arose of many different structures from a single physiological medium introducing a mechanism to select particles and shape to take into account this diversity. Obviously it has a character speculative. Measures can be done refuting or correcting the assumptions opening new questions about the role played by ionic currents on developmental processes. Unfortunately we have no idea to define the form the electrified surface is conformed to determine the shape of an organ. Even simple this model is not inconsistent with Wolper (1981 &1969) theory about positional information or Claudio Stern (1986) expectation's.

**FIGURE CAPTIONS**

Figure 1: A schematic representation of a section of the electrified cellular surface (figure 1a) showing the nucleus on the outmost cellular shell. In figure 1b a schematic representation of the different regions and surfaces near a single nucleus, which are relevant for the stochastic model (see text).

Figure 2: Potentiostatic plots of the current density vs. time obtained from the theory (Eq. 17). The curves are obtained for: D=1.x10$^{-5}$cm$^2$.s$^{-1}$, N=1.x10$^7$cm$^{-2}$, $n_0$k=1256.6 and $\alpha$=0. R assumes the following values: (a) R=1.58x10$^{-4}$cm, (b) R=3.16x10$^{-4}$cm, (c) R=1.26x10$^{-3}$cm and (d) R=3.16x10$^{-3}$cm.

Figure 3: Potentiostatic plots of the current density vs. time obtained from the theory (Eq. 18). D, N, $n_0$k and $\alpha$ are the same used into the curves of figure 2. R is fixed in R=3.16x10$^{-4}$cm whereas K assumes the values: (a) 1.x10$^{-11}$cm$^3$s$^{-1}$, (b) 1.x10$^{-9}$cm$^3$s$^{-1}$, (c) 1.x10$^{-8}$cm$^3$s$^{-1}$, (d) 1.x10$^{-7}$cm$^3$s$^{-1}$ and (e) 5.x10$^{-7}$cm$^3$s$^{-1}$.

Figure 4: Theoretical curves for $I/I_{max}$ vs. $t/t_{max}$ obtained form figure 2. The values of the parameters are the same as in figure 2 and the normalized current expressions are given by:

$$\frac{I}{I_{\max}} = C\left(1 + R.\left[\pi D t_{\max}\left( t \big/ t_{\max} \right)\right]^{-1/2}\right)\left(1 - \exp\left[-n_0 k t_{\max}\left( t \big/ t_{\max} \right) - \alpha k t_{\max}\left( t \big/ t_{\max} \right)^2\right]\right),$$

where $C$ is eq. (17) divided by $NbC_1$ when $t = t_{\max}$.

Figure 5: Theoretical curves for $I/I_{max}$ vs. $t/t_{max}$ obtained from figure 3. The value of the parameters is the same as in that figure. K change in each curve: (a) K=1.x10$^{-7}$cm$^3$s$^{-1}$, (b) K=1.x10$^{-8}$cm$^3$s$^{-1}$ and (c) K=1.x10$^{-9}$cm$^3$s$^{-1}$. We have not show the curves for K=5.x10$^{-7}$cm$^3$s$^{-1}$ and 1.x10$^{-11}$cm$^3$s$^{-1}$ as in figure 3. The curves obey equation (18) after the double normalization procedure. Full circle and full square curves stand for experimental results in electrolytic cells (see text).

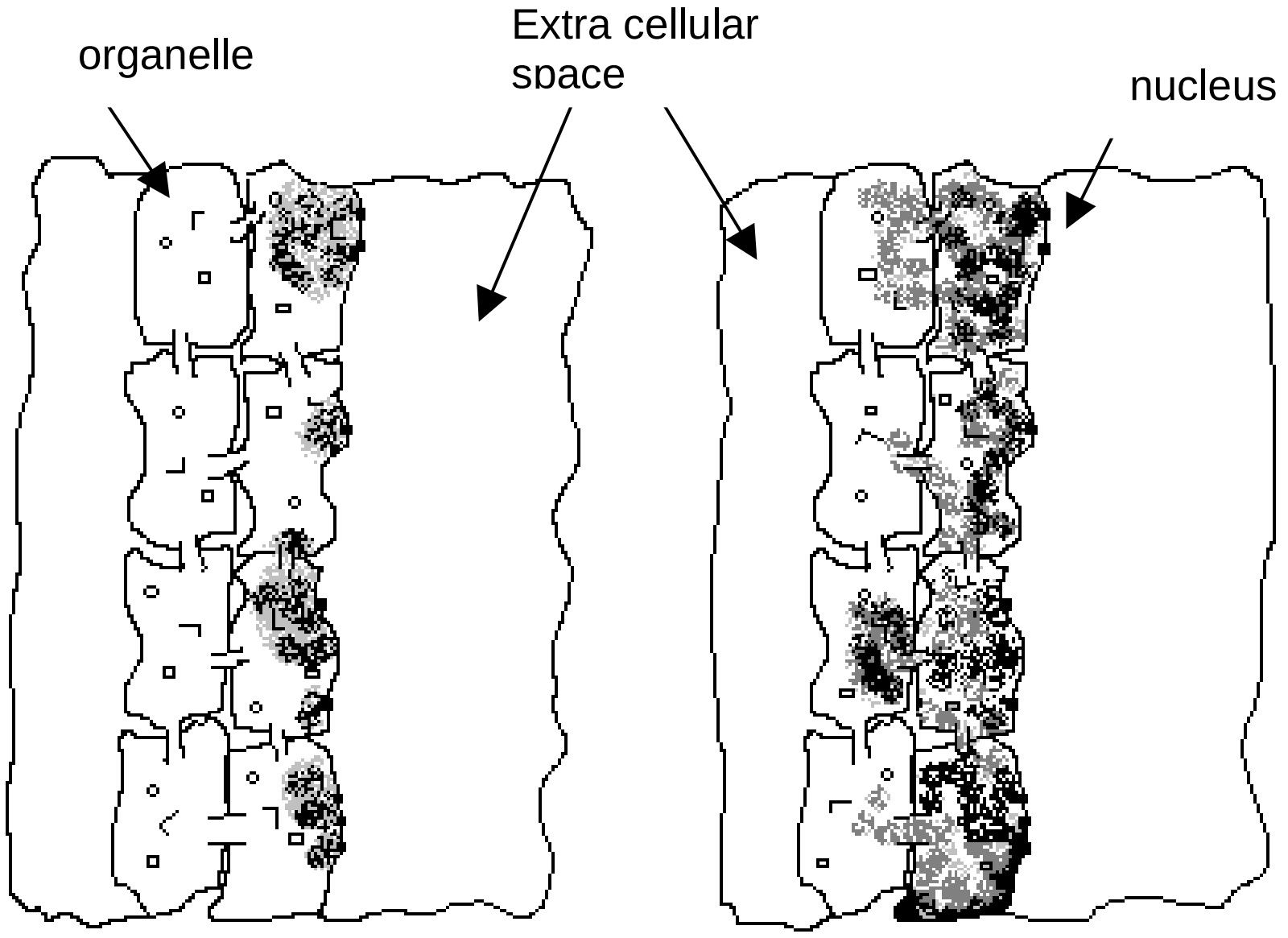

1.a

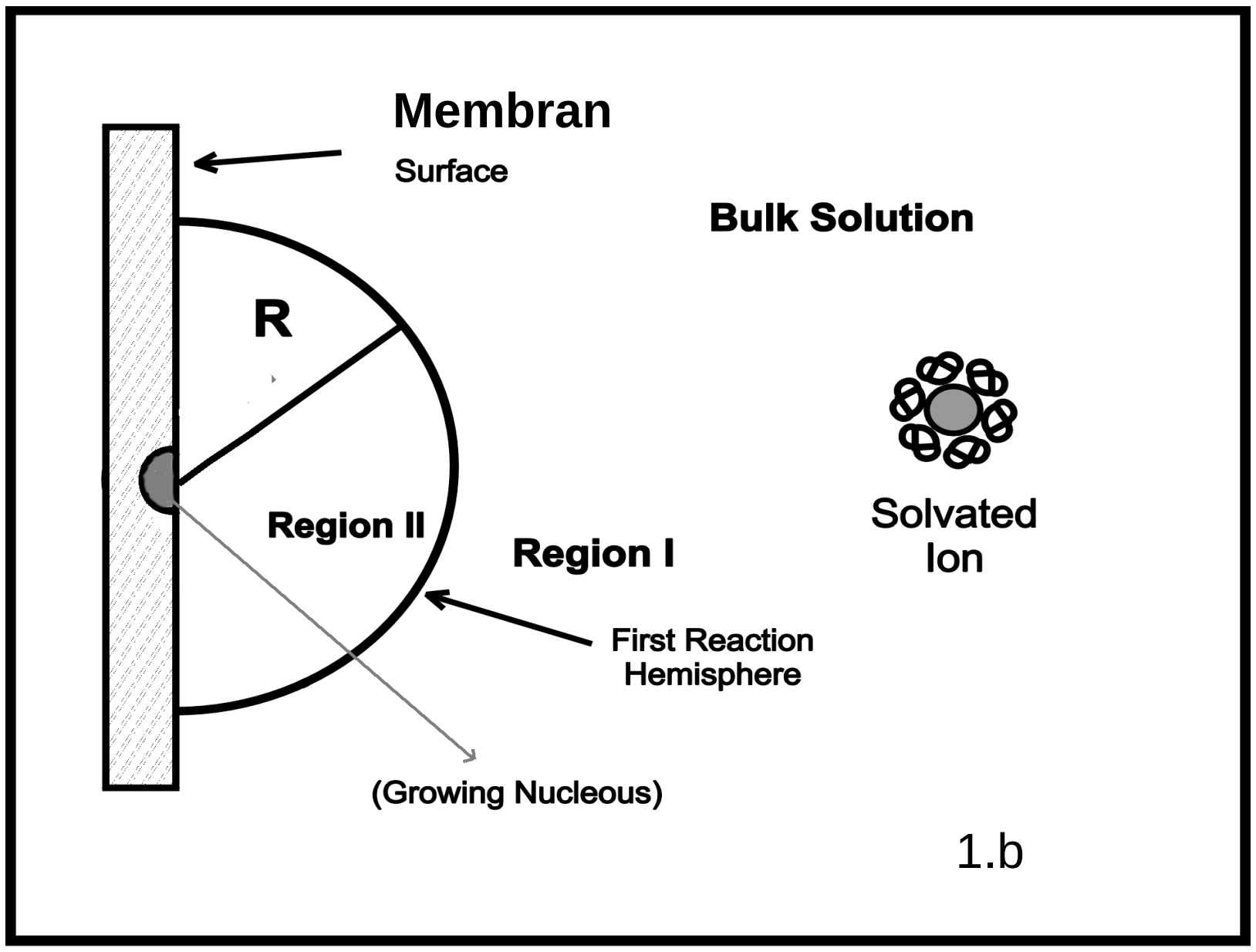

1.b

P.C.T.D'Ajello.... - Figure 1

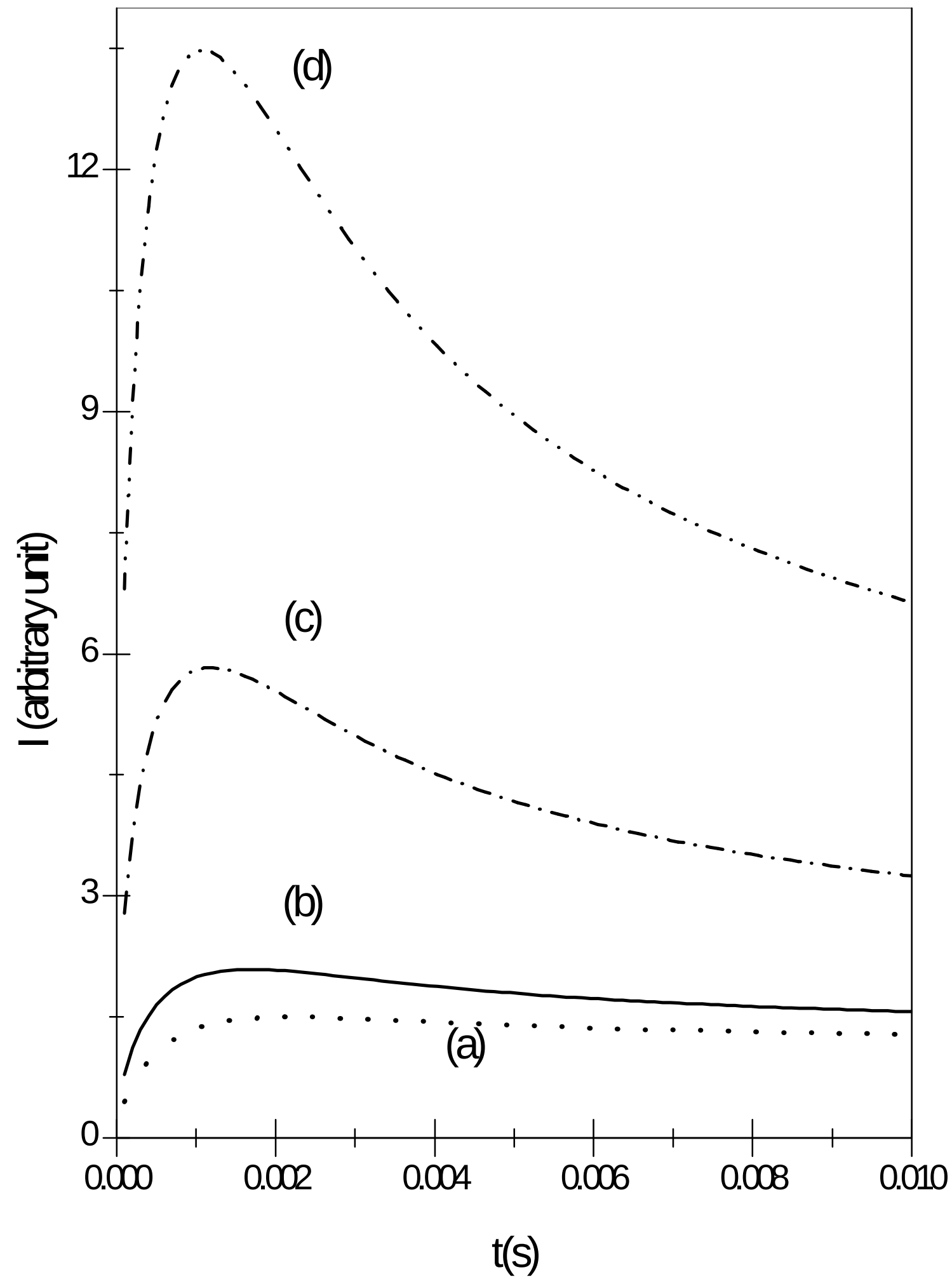

Figure 2 - D'Ajello et al

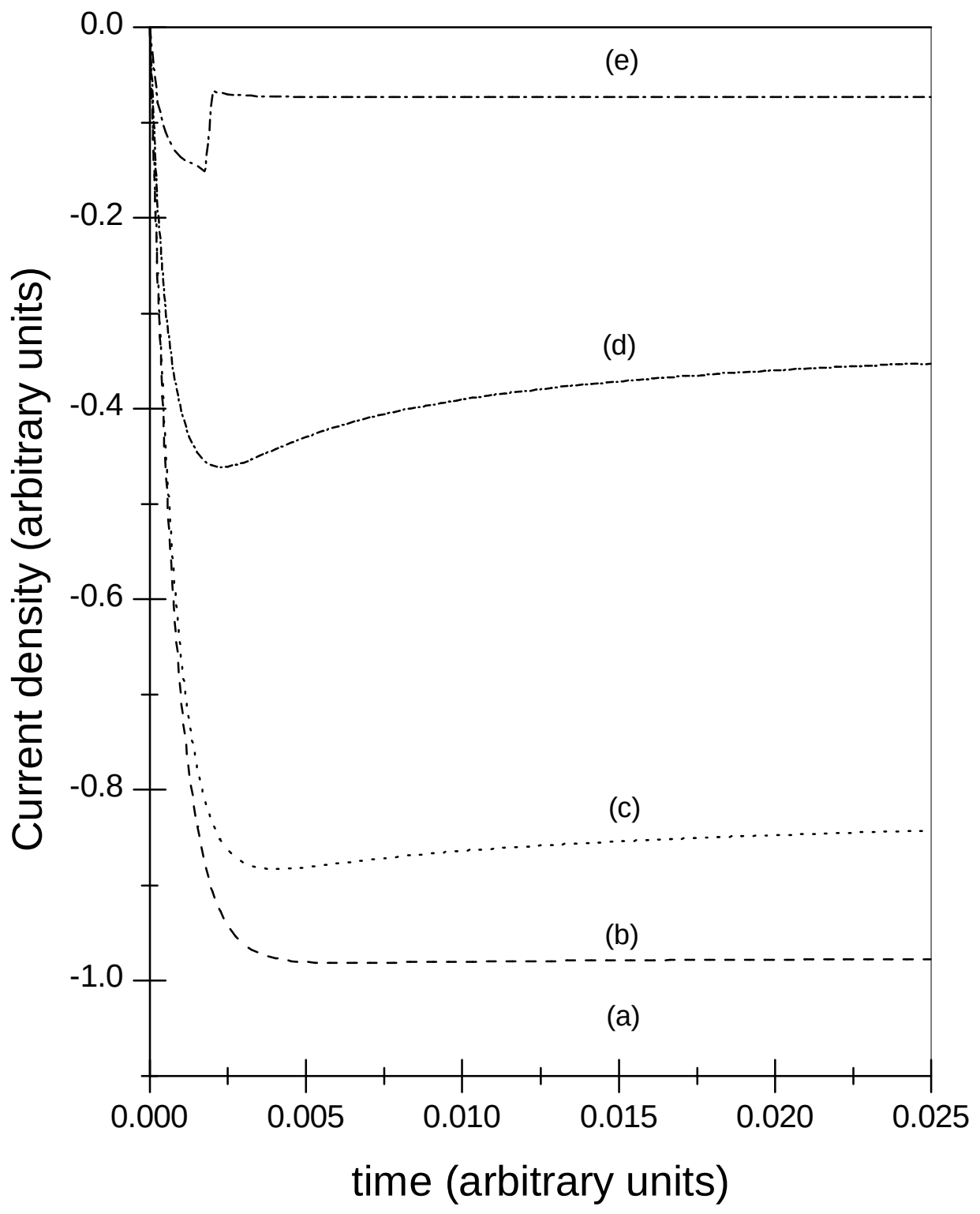

Figure 3- P.C.T. D'Ajello et al.

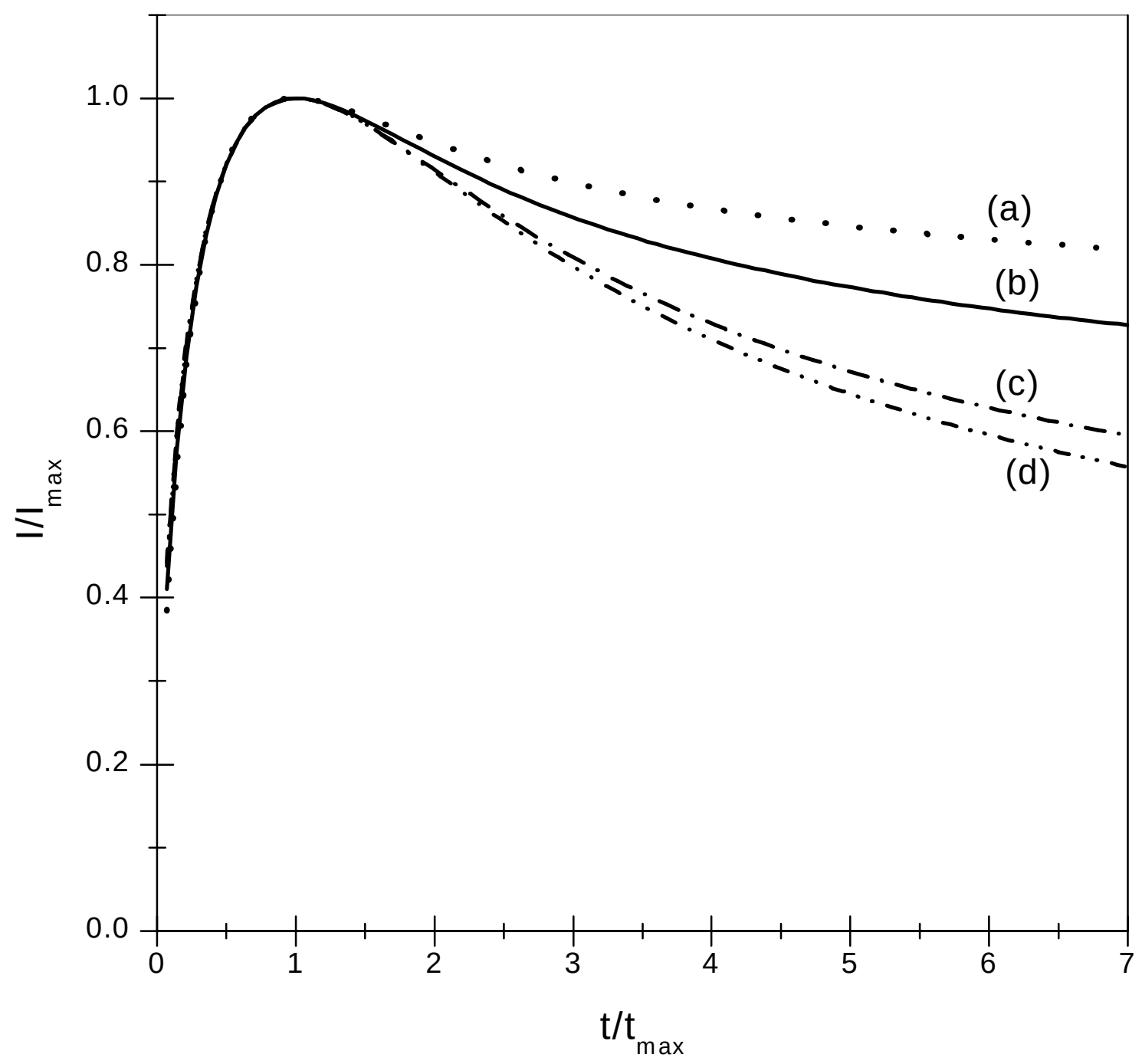

Figure 4 - D'Ajello et al

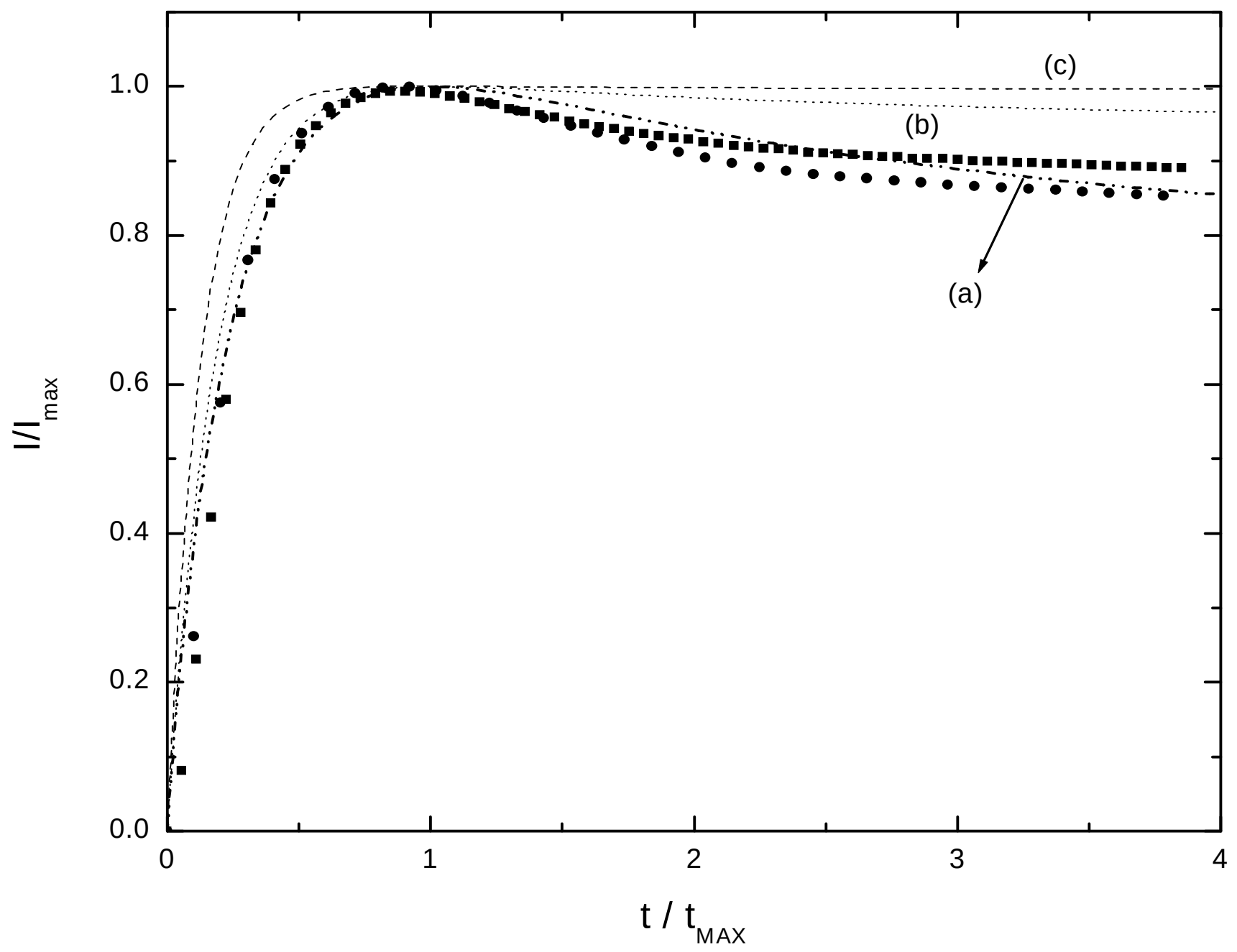

Figure 5 - P.C.T. D'Ajello et al.